\documentclass[useAMS,usegraphicx]{mn2e}

\title[Aspects of Galactic Accretion at Redshift 3.3]{Observational Aspects of Galactic Accretion  at Redshift 3.3\thanks{This paper includes data gathered with the 6.5 meter Magellan Telescopes located at Las Campanas Observatory, Chile.}}
\author[Michael Rauch et al.]{Michael Rauch,$^{1}$, George D. Becker,$^{2}$, Martin G. Haehnelt$^{3}$\\
$^{1}$Carnegie Observatories, 813 Santa Barbara Street, Pasadena, CA 91101, USA\\
$^{2}$Department of Physics \& Astronomy, University of California, Riverside, 900 University Ave, Riverside, CA, 92521, USA\\
$^{3}$Institute of Astronomy and Kavli Institute for Cosmology, Cambridge University, Madingley Road,  Cambridge CB30HA, UK\\
}
\begin{document}



\maketitle


\label{firstpage}

\begin{abstract} 
We investigate the origin of extragalactic continuum emission and its relation to the stellar population of a 
recently discovered peculiar z=3.344 Ly$\alpha$ emitter.
Based on an analysis of the broad-band colors and morphology we find further support
for the idea that the underlying galaxy is being fed by a large-scale (L$\ge 35$ kpc) accretion stream. Archival HST images
show small scale ($\sim 5$ kpc) tentacular filaments converging near a hot-spot of star-formation,  possibly fueled by gas falling in along the filaments. The spectral energy distribution of the tentacles is broadly compatible with either (1)  non-ionizing rest-frame far-UV continuum emission from stars formed in an 60 million-year-old starburst; (2) nebular 2-photon-continuum  radiation, arising from collisional excitation cooling, or  (3)  a recombination spectrum emitted by hydrogen fluorescing in response to ionizing radiation escaping from the galaxy. The latter possibility simultaneously accounts for the presence of asymmetric Ly$\alpha$ emission from the large-scale gaseous filament and the nebular continuum in the smaller-scale tentacles as caused by the escape of ionizing radiation
from the galaxy. Possible astrophysical explanations for the nature of the tentacles include: a galactic wind powered by the starburst; in-falling gas during cold accretion, or tails of interstellar medium dragged out of the galaxy by satellite halos that have plunged through the main halo. 
The possibility of detecting extragalactic 2-photon continuum emission in space-based, broad-band images suggests a tool for studying the gaseous environment of high redshift galaxies at much greater spatial detail than possible with Ly$\alpha$ or other resonance line emission. 
\end{abstract}


\section{Introduction}

\begin{figure}
\includegraphics[scale=.35,angle=0,keepaspectratio = true]{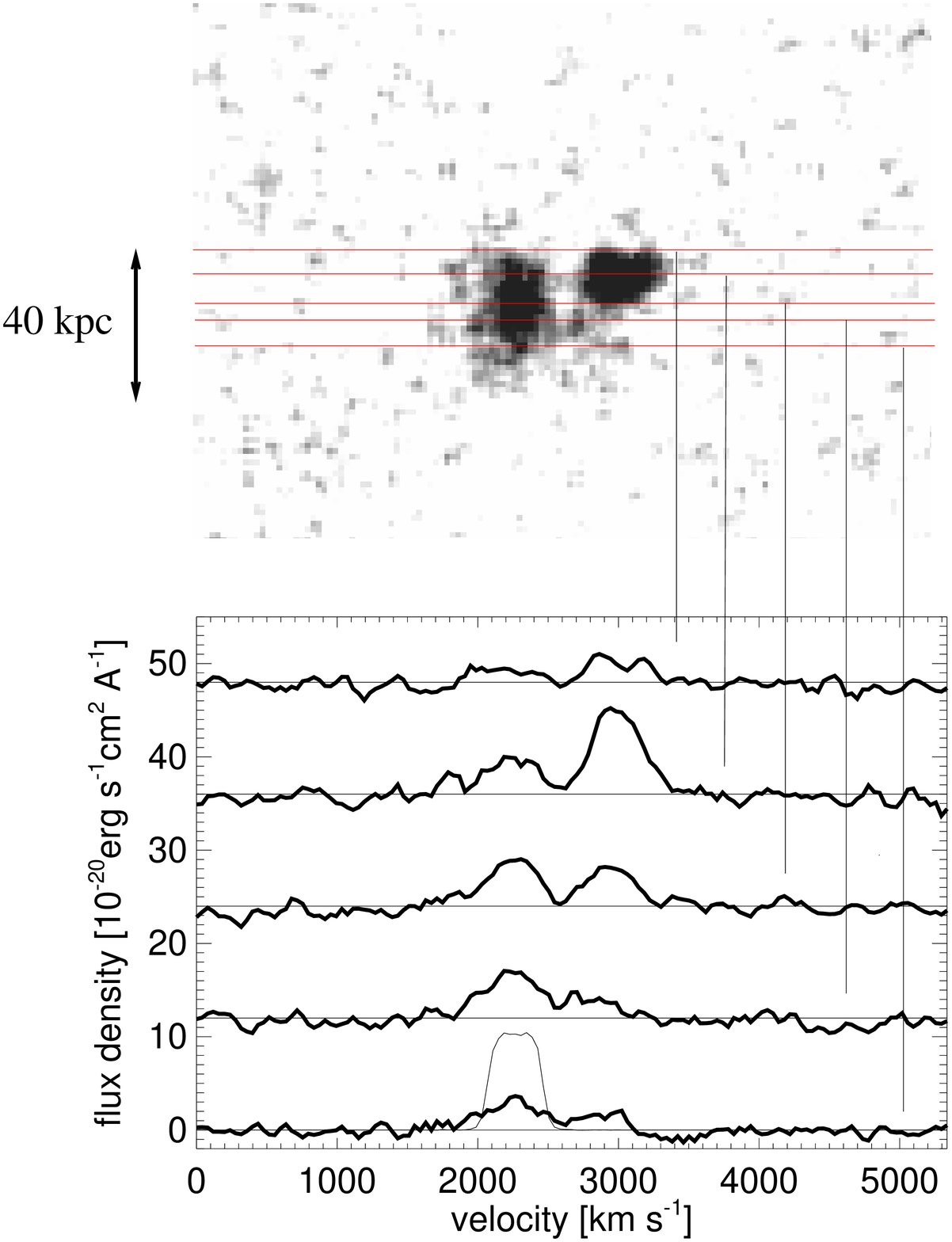}
\caption{ Spectral slices through the 2-dimensional spectrum of the Ly$\alpha$ emitting region. The dispersion runs from left to right (blue to red),
the spatial direction along the slit from bottom (South) to top (North). The solid curves in the bottom half of the plot are 3-pixel wide (0.56") spectra extracted along the 
parallel thin lines shown in the top panel. The top-most spectrum approximately corresponds to the position (along the slit) of the galactic continuum, which
cannot be seen here because of the DLA absorption trough.
The thin-line flat-topped profile in the bottom spectrum shows the spectral resolution profile
for a line source filling the slit (from a HeNeAr calibration lamp).\label{cuts}}
\end{figure}

Exchanges of gas between galaxies and the ambient intergalactic medium are thought to play an important role
in the formation of galaxies. Inflows of gas may be the dominant feeding mechanism for most galaxies through most of cosmic time. Outflows of some kind are thought to regulate the galactic baryon/metal budgets, galactic morphology, the shape of the luminosity function, and the enrichment of the IGM with metals. Intergalactic gas flows can be detected from the absorption lines they cause in the spectra of background QSOs or bright galaxies. Because of the sparse spatial sampling and the largely statistical nature of such constraints,  establishing a unique correspondence between absorption lines and galactic in- or outflows has remained difficult and highly model-dependent. An unambiguous understanding of the nature of intergalactic gas flows requires some form of direct imaging of the spatially extended emission from the galactic gaseous environment.

Taking 2-dimensional images of high redshift gas in emission is most likely to succeed by observing the redshifted HI Ly$\alpha$ emission line, powered by a strong source of ionizing radiation. Thus, most of the detections of extended emission so far are due to fluorescence of Ly$\alpha$ in response to ionizing photons in the vicinity of an AGN (e.g., Fynbo et al 1999; Bunker et al 2003; Francis \& Bland-Hawthorn 2004; 
Weidinger et al 2004; Cantalupo et al 2005,2007; Kollmeier et al 2010; Adelberger et al 2006;, Hennawi et al 2009; Trainor \& Steidel 2013; Martin et al 2014a,b, 2015). Extended Ly$\alpha$ emission is also seen among emitters  detected in narrow-band imaging surveys without direct reference to the presence of an AGN. At the bright end of the luminosity distribution of these objects, the so-called “Ly$\alpha$ blobs” are likely to reflect an inhomogeneous population of gaseous halos lit up by, again, AGN (e.g., Overzier et al 2013, Prescott et al 2015, and references therein), strong star-formation (e.g., Cen \& Zheng 2013; Ouchi et al 2013; Zabl et al 2015) cooling of gas falling into gravitational potential wells, (e.g., Haiman, Spaans \& Quataert 2000; Fardal et al 2001) and possibly other feedback processes.
The simultaneous association of these processes with the formation of massive halos so far has made it difficult to disentangle them observationally.
Studying the gas/galaxy interface of a more typical, lower mass galaxy one would expect star-formation to become the dominant source of Ly$\alpha$ emission, with the main difficulty shifting to being able to detect extended emission related to star formation rates on the
order of a few solar masses per year. For a $z\sim 3$ galaxy, a star-formation rate of 1 $M_{\odot}$yr$^{-1}$, translates into a total, unattenuated Ly$\alpha$ flux (predicted for case B conditions) of $1.4\times 10^{-17}$ erg s$^{-1}$ cm$^{-2}$, about an order of magnitude fainter than Ly$\alpha$ blobs typically uncovered by narrow band surveys (e.g., Matsuda et al 2004).
The higher detection sensitivity required to see extended halos at such fluxes can be achieved through long exposures and a spectroscopic search strategy as opposed to a traditional narrowband filter
survey. Trying to detect Ly$\alpha$ sources (observed through a blindly positioned spectrograph slit) in a 2-d spectrum considerably reduces the sky background noise  by a factor $\sqrt{\Delta \lambda_{\rm NB}/\Delta \lambda_{\rm spec}}$ where $\Delta \lambda_{\rm NB}$ is the width of a typical narrow band filter, and $\Delta \lambda_{\rm spec}$ is the spectral resolution element of a typical low resolution spectrograph. A similar suppression can be achieved by applying a judiciously positioned “slice” in wavelength space a posteriori through the datacube recorded with an integral field unit (IFU)). 
Such surveys have reached surface brightness limits of $8\times10^{-20}$ erg cm$^{-2}$ s$^{-1}$ arcsec$^{-2}$ per arcsec$^2$ aperture (Rauch et al 2008). At this level many $z\sim 3$ galaxies {\it individually} show extended Ly$\alpha$ emission, even with star-formation rates less than 1 $M_{\odot}$yr$^{-1}$ (Rauch et al 2008). The brighter ones of those objects exhibit surface brightness profiles best described as a relatively compact core with broad wings (Rauch et al 2013a, fig. 6), which can be reproduced by simple models of HI halos where Ly$\alpha$ photons propagate from a central source through an optically thick HI coccoon  (e.g., Dijkstra et al 2006; Barnes \& Haehnelt 2010; Rauch et al 2013a, fig. B1). Studies stacking narrowband images of many halos of individually lower S/N (e.g., Hayashino et al 2004; Steidel et al 2012, Feldmeier et al 2013; Momose et al 2014) confirm this finding of extended halos in an average sense, but suggest that  the ensemble properties of the extended halos are sensitive to the sample selection criteria (e.g., broad band-selected vs. narrowband selected), and to the redshift.

Among the individually extended halos found in our earlier searches, there is a subset of widely extended (several tens of kpc wide),
Ly$\alpha$ halos, whose surface brightness profiles do not exhibit to the spatial symmetry or the rapid drop
commonly found among Ly$\alpha$ emitting galaxies. These objects include halos with vaguely filamentary extensions, apparently corresponding to in-falling gas and satellites (Rauch et al 2011, 2013a, 2014), and in one case, a candidate for star-formation triggered by external AGN feedback (Rauch 2013b). As these cases are rare relative to the more common spatially symmetric and compact Ly$\alpha$ emitters it appears that they represent phases in the formation of galactic halos, where short-lived phenomena
like starbursts or AGN activity briefly light up the distant gaseous surroundings of high z galaxies. 

\smallskip

In this paper we revisit one of these objects, the peculiar extended and asymmetric z=3.344 Ly$\alpha$ emitter related to a galaxy at position 03:32:38.815, -27:46:14.34 (2000), known as UDF ACS 07675, or GOODS-CDFS-MUSIC 11517 (e.g., Grazian et al 2005). Our earlier analysis of this object (Rauch et al 2011, paper I), based on the discovery spectrum taken with LDSS3 on the Magellan-II telescope, showed an area of complex, anisotropic Ly$\alpha$ emission continuously traceable out to about 35 $h_{70}^{-1}$ proper kpc to the South of the galaxy. 
We interpreted the blueshift of the main, linear stretch of Ly$\alpha$ emission with respect to the damped Ly$\alpha$ absorption  (DLA)
by the galaxy, and the fact that the DLA partly absorbs the emission, as evidence for backside infall of a filamentary structure of gas onto the galaxy. 
The fact that the filament is hardly tilted or distorted in velocity space suggests a velocity that is more or less constant when approaching the galaxy, which
is predicted by some theoretical studies of streams in the cold accretion scenario (e.g., Goerdt \& Ceverino 2015, and references therein).
The large extent and simple velocity structure of the filament, together with its extreme asymmetry (it is visible
only to one side of the galaxy) suggested that it may be gas fluorescing in response to ionizing radiation escaping anisotropically from
the galaxy itself.

The earlier paper was concerned mainly with the Ly$\alpha$ emission and the gaseous environment. 
Here we extend our analysis to include a closer look at the observational properties of the underlying galaxy, as they appear in
the broad band images taken with the HST ACS camera by the  Hubble Ultra Deep Field project (HUDF; Beckwith et al 2006).
The present study will use these imaging data to shed light on our original interpretation of the nature of the Ly$\alpha$ emitter, and constrain the properties of the underlying flows of gas and stars. We will argue that some of the broad band emission seen in the HST images can be interpreted as extragalactic nebular continuum emission, and discuss several physical explanations, in particular the
fluorescence of gaseous in- and outflows illuminated by ionizing photons escaping from the galaxy.

\section{Observations}

\subsection{Appearance of the Ly$\alpha$ emission}

Fig. \ref{cuts} shows the 2-dimensional LDSS3 spectrum around the Ly$\alpha$ emission line, together with several 1-dimensional, 3-pixel-wide ($3\times0.188"$) extracted spectral cuts. Two main regions of emission can be seen, a more compact one to the right (red) side that we called the "Red Core" in 
paper I, which, aside from its spatial asymmetry, looks similar to a standard Ly$\alpha$ halo (e.g., Rauch et al 2013a). To the left and blueward of the Red Core extends a triangular region (the "Blue Fan"). The blue end of the Blue Fan is occupied by a filamentary ridge of emission (the "Filament"; see also \ref{fig2}). In the original analysis, the filamentary structure appeared
to consist of two well separated ridges which could be traced out to several tens of kpc, but re-reducing the data with an improved wavelength registration now shows these ridges closer together and merging into one sharper structure beyond 7 kpc away from the galaxy. The total fluxes in the filament and the red core are comparable. At distances along the slit beyond 10 kpc from the galaxy, the flux density in the filament begins to dominate over
that of the red core. The spectral cuts show that - except perhaps for the central cut -  the filament and the red core are narrower in velocity
space than predicted for a light source filling the slit (see the resolution profile near the bottom cut in the figure), suggesting that they
are smaller in the E-W direction than the 2" slit-width.

\subsection{Ly$\alpha$ emission and the underlying galaxy}

\begin{figure*}
\includegraphics[scale=.40,angle=0,keepaspectratio = true]{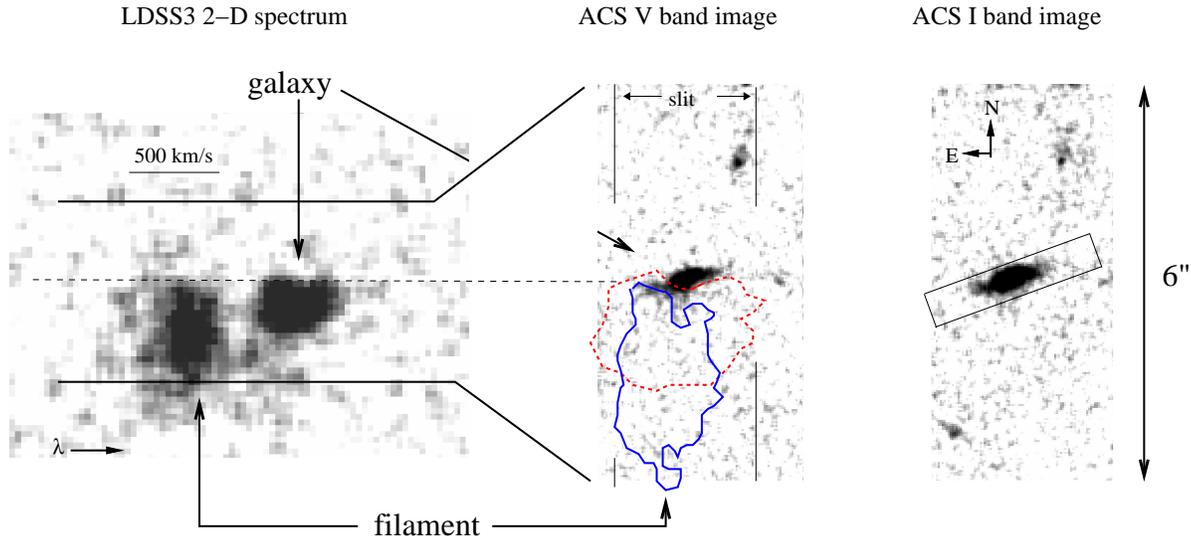}
\caption{Left panel: Two-dimensional spectrum of the Ly$\alpha$ emission line, with the spectral dispersion horizontally, and the spatial direction along the slit vertically.  The spatial position of the continuum is indicated by the thin dashed horizontal line. As argued in paper I, the triangular emission fan encompasses a compact, red-dominant Ly$\alpha$ emission peak presumably arising in the direct vicinity of the galaxy, and a blue-shifted,  'filament' (or multiple unresolved filaments) of emission that is far more spatially extended along the slit. Middle panel: F606W image of UDF ACS 07675. Note the change in spatial scale between spectrum and image. Faint emission in both bands extends South (bottom)
and West (right) of the galaxy.  The two thin vertical lines indicate the slit edges. The thin solid (blue) contour in the middle panel delineates approximately the $5\times 10^{-20}$ erg cm$^{-2}$s$^{-1}\AA^{-1}$ flux density level of the spectrum of the 'filament'; the dotted (red) contour corresponds to the redder ('red core') emission region.
Both were scaled and overplotted on the image to demonstrate the relative spatial extents of the Ly$\alpha$ and the continuum emission in the direction along the slit (N-S).
This is for illustrative purposes only, as the E-W width and alignment
between the flux density contours (which are observed in wavelength space) and the broad band image is uncertain beyond the requirement
that both have to have arisen from a region within the projected footprint of the spectrograph slit. 
The rightmost panel shows the F775W I band image with the rectangular extraction aperture used for the spatial cut in surface brightness shown below in fig. \ref{colors}.\label{fig2}}
\end{figure*}

Further constraints on the nature of the emitter can be obtained from the publicly available HST ACS images recorded by the HUDF project
(Beckwith et al 2006). Fig. \ref{fig2} shows how the Ly$\alpha$ emission and the underlying galaxy relate to each other. The 2-d spectrum of the Ly$\alpha$ emission line is given
in the leftmost panel, and  the ACS V (F606W) and I (F775W) band images of the underlying galaxy
UDF-ACS-07675 in the central and right panels, respectively. For better visibility, the broad band images are somewhat enlarged compared to the spectrum. Spectrum and images are lined up in the direction along the slit (N-S) to make the faint spectral continuum trace (not visible here but
indicated by a dashed line) coincide with the centroid of the emission in the V-band. For illustrative purposes, the outlines of the blue (solid) and red (dashed) Ly$\alpha$ emission components
are arbitrarily shifted onto and overplotted on the V band image, with the contours representing a  Ly$\alpha$ flux density  at approximately the $5\times 10^{-20}$ erg cm$^{-2}$s$^{-1}\AA^{-1}$ level. Because the contours are taken from a spectrum, they are a combination of spatial and velocity structure, and thus the precise spatial alignment with the image along the dispersion direction and its spatial width in the dispersion direction is uncertain.

It can be seen that the two Ly$\alpha$ components occur only to the South of the galaxy.
The V band image (center of fig.\ref{fig2}; for a detailed view see fig.\ref{blowup}) shows tentacles of emission emerging from the galaxy in the general direction of the Ly$\alpha$ emission, and an additional tail with a distinct head extending
about 1.5" to the right (i.e., West). The flux in the tentacles is statistically significant at the 7.7 (V band), 4.7 (I band) and 5.5 (z band) $\sigma$ levels, and the
flux in the tail is a 4.9 (V band) and 4.3 (I band) $\sigma$ detection.

\subsection{Appearance of the immediately galactic neighbourhood}

\begin{figure*}
\includegraphics[scale=.45,angle=0,keepaspectratio = true]{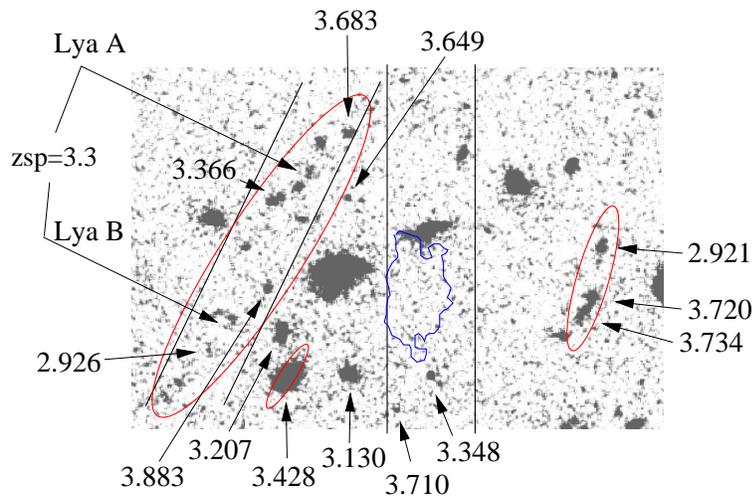}
\caption{$12.6" \times 8.7"$ HUDF V band image of the larger scale environment of the main Ly$\alpha$ emitting galaxy. North is up and East to the left. The position of the two slits are shown. The positions (along
the tilted slit) of two new Ly$\alpha$ emitting spots detected at the same redshift as the main Ly$\alpha$ emitting galaxy (z=3.344) are given as "LyA" and "LyB". The other numbers are photometric redshifts
(from Coe et al 2006) for objects with $z = 3.344\pm 0.5$. The red ellipses mark structures tentatively suggestive of alignment with the central galaxy and its Ly$\alpha$ emitting "filament" (thin blue contour).
\label{largescale}}
\end{figure*}

An image of the larger scale environment of the galaxy is given in Fig.\ref{largescale}. A $12.6" \times 8.7"$ ($94\times 65$ proper kpc)
HUDF V band image shows the main z=3.344 galaxy  close to the center. The vertical straight lines are the outlines of the N-S oriented slit used to obtain the spectrum in figs. \ref{cuts} and \ref{fig2}. The tilted straight lines
show a second slit position that was used for auxiliary spectroscopy to probe the redshifts of an apparent filament of galaxies. This second spectrum, with a total exposure time of 66,000s is not shown here, as it is of much poorer quality than the main spectrum. It nevertheless shows  two spots with Ly$\alpha$ line emission at virtually the same redshift as the main Ly$\alpha$ emission complex. The spatial positions along the slit of these two emission spots are denoted by "LyA" and "LyB", and
by two arrows marked with "zsp=3.3 Ly$\alpha$". The remaining numbers in the image are photometric redshifts (from Coe et al 2006) within $\Delta z = \pm 0.5$ of the spectroscopic Ly$\alpha$ emission redshift 3.344 of the main galaxy. At the very faint magnitudes of these galaxies, the uncertainty of the published redshifts in this sample is large enough that all of them
could plausibly be residing at the same redshift. In addition, the morphological similarity of several objects to the ones having  spectroscopic Ly$\alpha$ redshifts, or to ones having photometric
redshifts closer to the spectroscopic one 
suggests that we treat these galaxies as candidate redshift=3.3 objects, although we should
expect false positives and negatives in this sample.
The linear grouping of the spectroscopic Ly$\alpha$ A and B z=3.3 redshifts with at least four and possibly more galaxies with similar redshifts (z=2.926, 3.883, 3.366, 3.683) situated under the second slit, and the existence of another linear group to the West (z=3.734, 3.720, 2.921) are suggestive of some larger scale structure coarsely aligned with the NNW-SSE opening cone of the tentacles of the main galaxy. The picture is made more complicated by the presence of several other candidates near the S bottom of the image (z=3.207, 3.428, 3.130, 3.710, 3.348), of which the brightest one (with z=3.428), consists of three knots of emission also aligned in the general direction toward the main galaxy. This object has a star formation rate of 43.9 $M_{\odot}$yr$^{-1}$ (Xue et al 2010), about ten times the value of the main galaxy.  It is about 3.8" away from the edge of the slit, but it could
conceivably have contributed, perhaps through fluorescence, to the Ly$\alpha$ emission covered by the slit.

\subsection{Detailed properties of the galaxy}

\begin{figure*}
\includegraphics[scale=.35,angle=0,keepaspectratio = true]{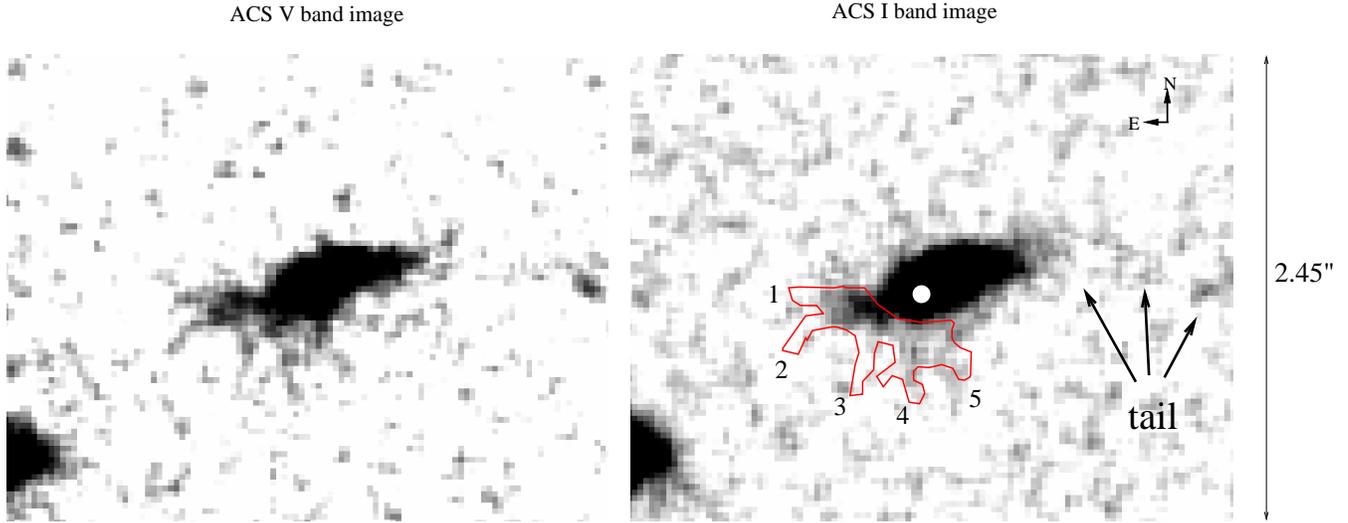}
\caption{Enlargement of the V and I band images, showing several tentacles of emission to the immediate bottom-left (South-East), and a longer tail extending for about $\sim 1.5"$ (11.5 kpc) to the (right) West, ending in a faint knot at the right edge of the images. The white spot in the right panel shows the position of peak V band emission.
The contours in the I band image give the outline of the tentacular V-band structures, to aid cross-identification with the I band image. The presence
of common emission features in both bands (the knot between 1 and 2, and the filaments 4 and 5; and the long tail) suggests the presence of continuum emission
in these structures.  \label{blowup}}
\end{figure*}

Closer inspection of the galaxy itself in HST broad band images reveals a number of additional interesting clues to the nature
of this object.
An enlarged image of the galaxy  is given in fig.\ref{blowup}.  The V-band image (left panel) shows
at least five tentacular features stretching SSE between 4 and 6 kpc from the center of emission (indicated in the I band image (right panel) as a white dot). A crude, hand-drawn contour outline
of the tentacles in the V band image, overlaid on the I band image, shows some of the appendages (the knot between 1 and 2, and the features 4 and 5, and the "tail" to the right) to be present in the I band as well.  The presence of emission in both filters implies that we are seeing some form of continuum radiation. The tail to the right looks like a typical tidal feature, e. g., a dwarf galaxy that
may have passed through the main object, with its flux peaking at the end of the structure. The tentacles to the South look more unusual, on account of their number, their similar length and surface brightness, and the fact that their flux is not peaked at the end of the tentacles, as one might naively expect if they were a swarm of tidal stellar features. 

\begin{figure*}
\includegraphics[scale=0.49,angle=0,keepaspectratio = true]{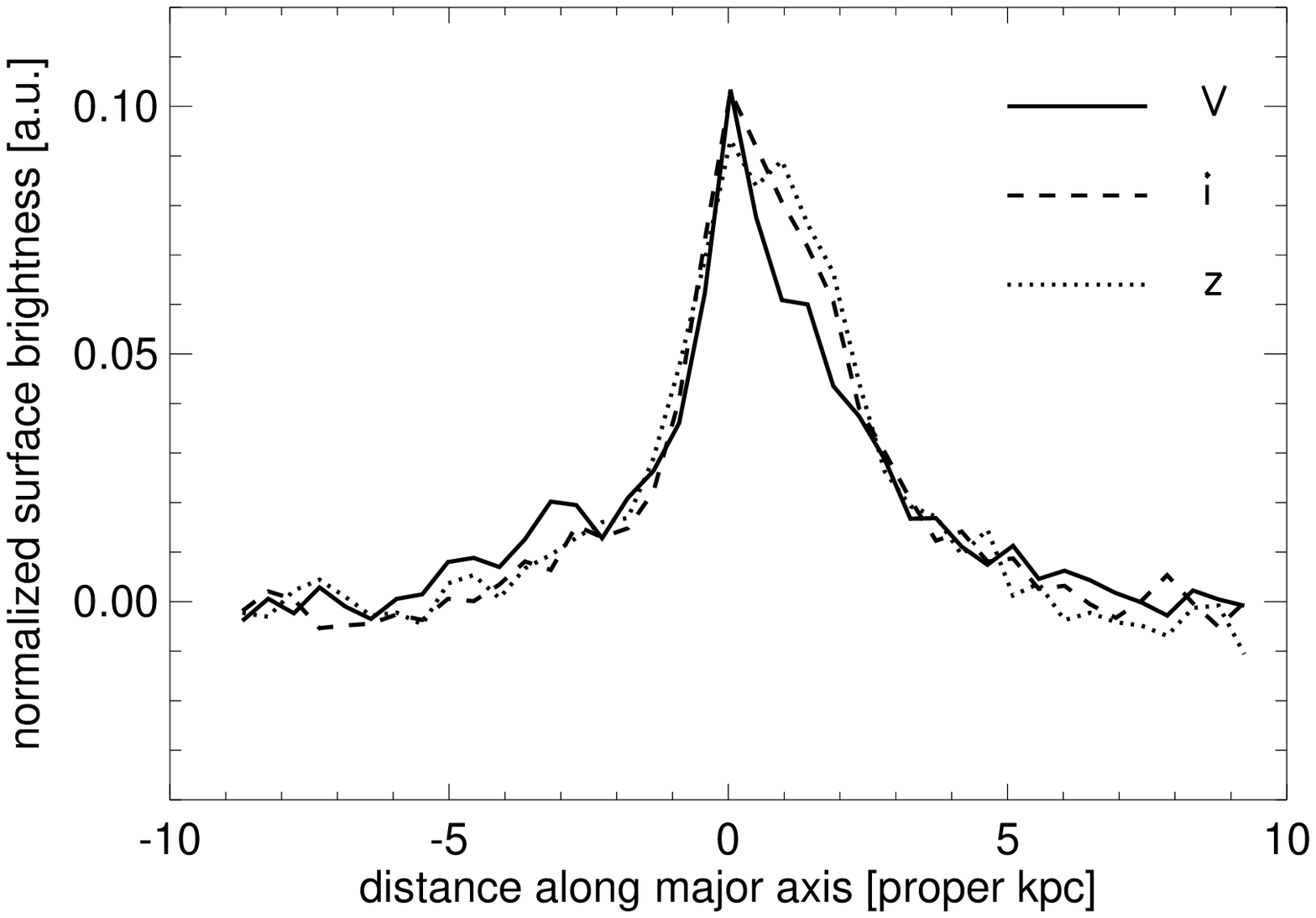}
\includegraphics[scale=0.49,angle=0,keepaspectratio = true]{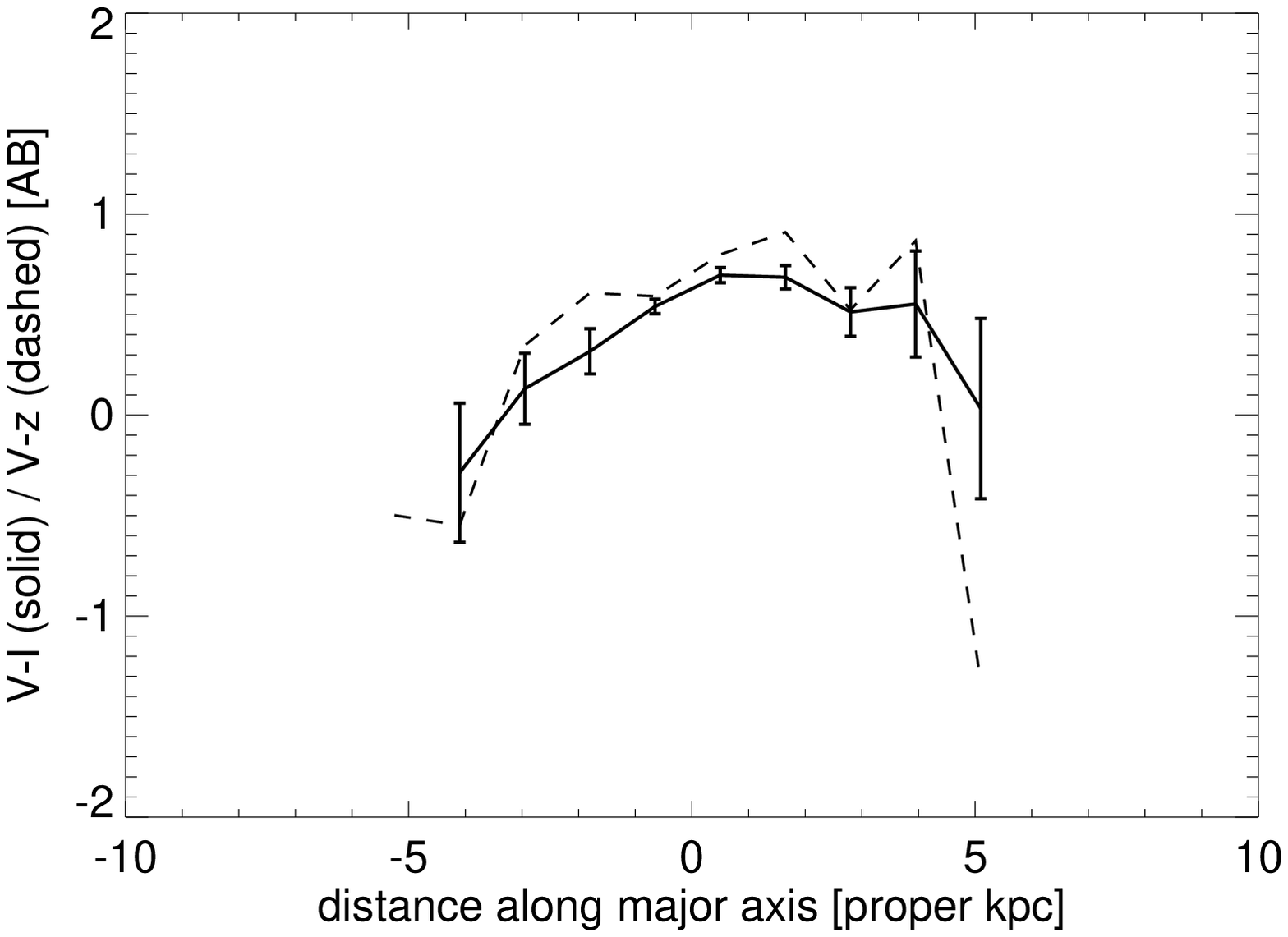}
\caption{Relative surface brightness per pixel  along a direction aligned with the approximate position angle of -20 degrees of the
galaxy. The flux was determined by adding up the pixel fluxes perpendicular to and across a width of 0.57" in an extraction window aligned with this direction. The origin of the coordinates was arbitrarily placed near the brightness peak. Left panel: normalized surface brightness in three bands; right panel: spatially smoothed V-I (solid) and V-z (dashed) colours
versus distance along the cut. To avoid confusion, only  the V-I errors are plotted.\label{colors}}
\end{figure*}

The left panel of fig. \ref{colors} shows the relative surface brightness along a linear cut with a position angle of -20 degrees, approximately aligned with
the direction of largest extent of the galaxy. The flux was determined by adding up the pixel fluxes perpendicular to and across a width of 0.57" in a rectangular extraction window aligned with this direction (its outline is shown in the rightmost panel of fig. \ref{fig2}). It can be seen that the flux is sharply peaked and increasingly asymmetric going blueward from z via I to the V band,  giving the galaxy its distinct tadpole appearance. Aside from the peak, the bluest colors occur between -6 and -2 kpc, ie., they arise in the small scale tentacles.

\section{Physical origin of the filamentary continuum emission}

The origin of the continuum radiation detected in the broad images of the tentacular region is not immediately clear. 
The standard expectation, that this is just another case of  stellar continuum emission, is not necessarily the only one. Given the location of  the galaxy in a large and obviously gas-rich Ly$\alpha$ emitting environment, and the unusual shape and relatively homogeneous surface brightness of the tentacles we may be seeing an extragalactic incidence of nebular continuum emission of some sort. Thus,  we shall discuss three main alternative astrophysical sources for the observed continuum radiation: (1) a stellar continuum; (2) collisional excitation 2-photon continuum; and (3) a fluorescent recombination spectrum, emitted in response to ionizing radiation from stars in the galaxy.

\begin{figure*}
\includegraphics[scale=.53,angle=0,keepaspectratio = true]{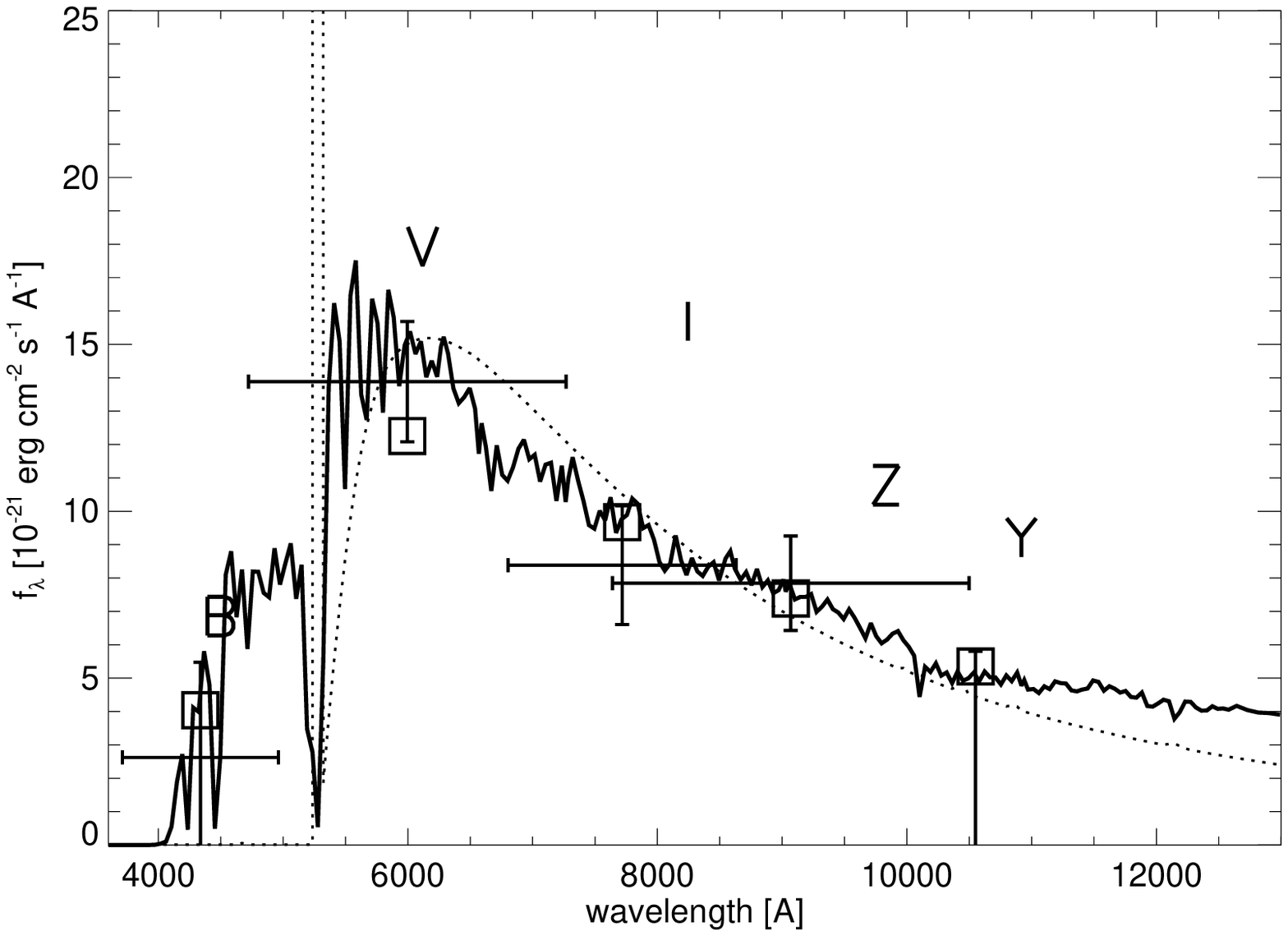}
\includegraphics[scale=.53,angle=0,keepaspectratio = true]{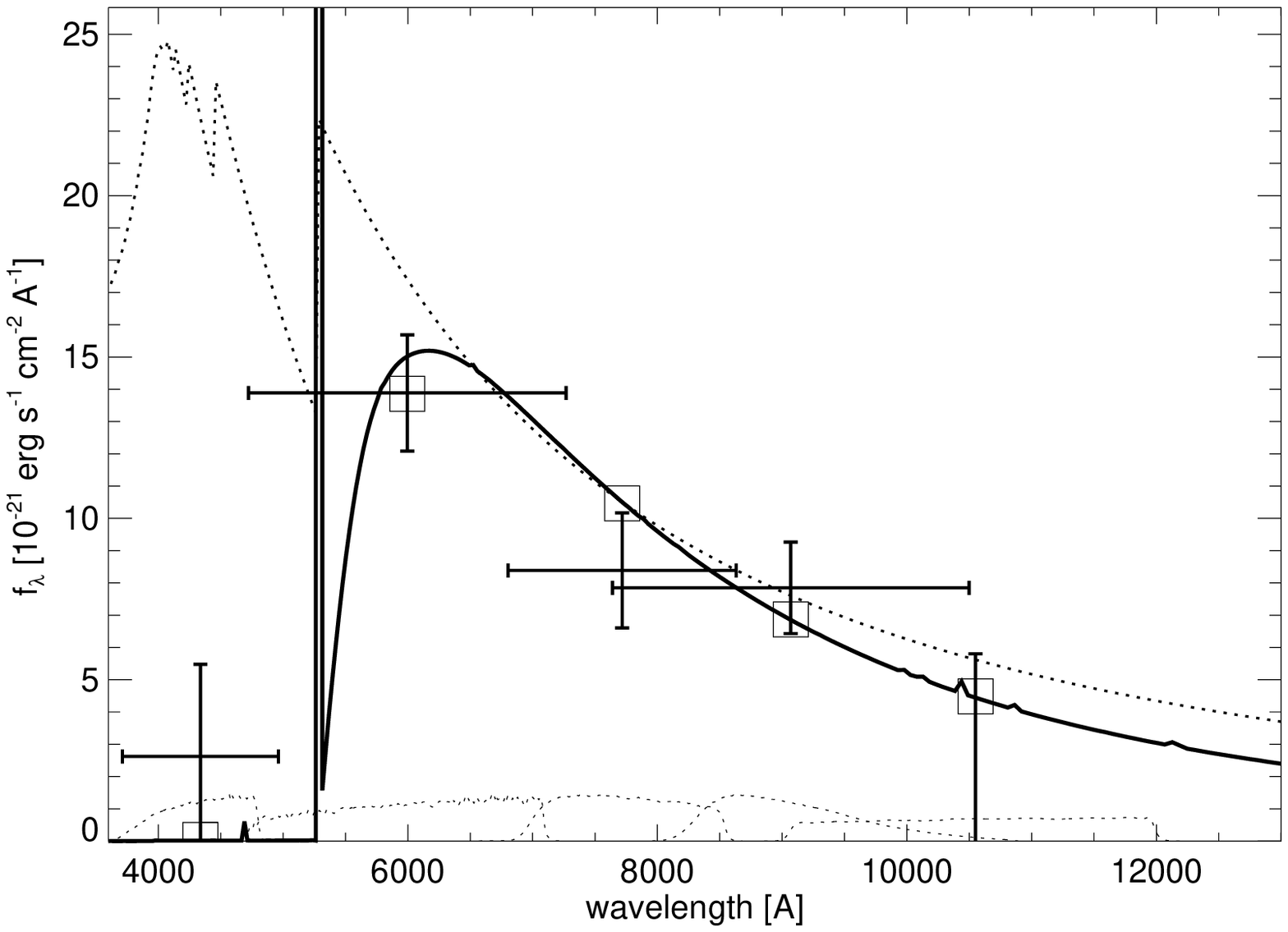}
\includegraphics[scale=.53,angle=0,keepaspectratio = true]{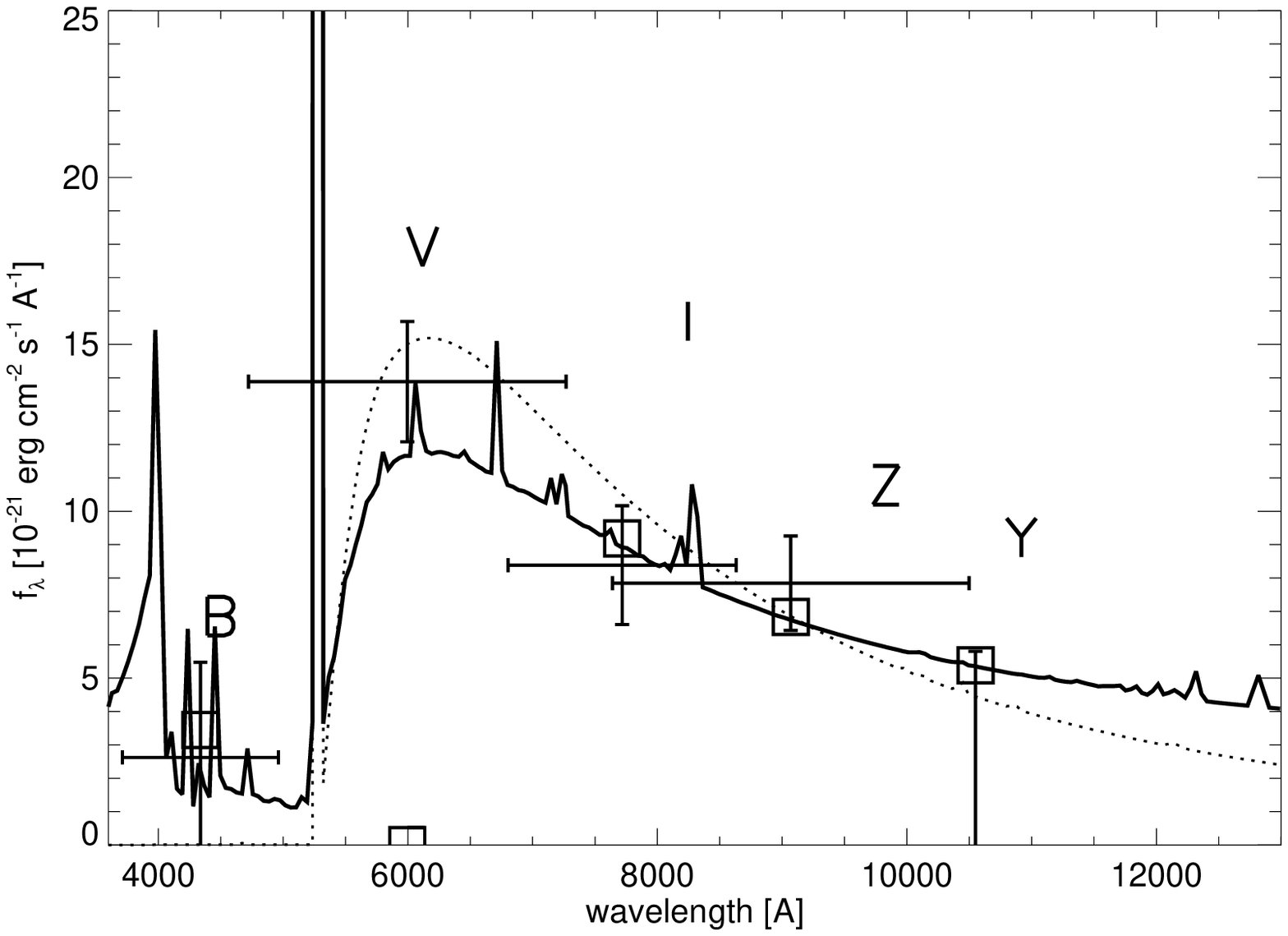}
\caption{Spectra of three alternative models for the broad band colours of the tentacles. The models are trying to reproduce the emission as coming from either: (1)  a starburst
stellar continuum (top), (2) collisionally excited cooling radiation (center) or (3) a fluorescent recombination spectrum of gas exposed to an ionizing stellar continuum (bottom). The data points with  error bars represent the observed fluxes in the various bands from within the contour shown in the right panel of fig.\ref{blowup}. The little square boxes give the integrated fluxes passing through the same filters for the model spectra. The top panel shows the spectrum of a $6\times10^7$ yr old starburst from STARBURST99 (solid line). The central panel shows the spectrum of collisionally excited gas  from a Cloudy model described in the text (solid line;  also over-plotted for comparison as a dotted line in the left and right panels). The dotted line in the central panel shows a continuum of the form $f_{\lambda}\propto \lambda^{-\beta}$, with $\beta = 2$ for illustration.
The curves at the bottom of the central panel are of the transmission curves of the HST filters (arbitrary units). 
The bottom panel shows a fluorescent recombination spectrum produced by ionizing radiation from a starburst.
All spectra were attenuated by the appropriate Ly$\alpha$ forest absorption (Inoue et al 2014).  In the central panel the V band was fit separately to account for the unknown
Ly$\alpha$ escape fraction. In the bottom panel, no fit was attempted for the V band. \label{tentacont}}
\end{figure*}

\subsection{Stellar continuum radiation}

The broad  band colours of the tentacle region can be well represented by the continuum emission from a conventional stellar population.  
Fitting STARBURST99 (Leitherer et al 1999) galactic spectra of instantaneous starbursts of different ages to the observed HST B, V, I, z and Y band images (with the additional Y band taken from the CANDELS survey; Koekemoer et al., 2011), the best fit (with a  reduced
$\chi^2_{\nu} = 0.85$) was achieved with a stellar population with age of $6\times10^7$ yr (top panel in fig.\ref{tentacont}).

\subsection{Collisionally excited 2-photon continuum}

Alternatively, the continuum emission could represent the 2-photon continuum emitted in competition
to Ly$\alpha$ line radiation (e.g., Dijkstra 2009, and references therein) during either recombination or collisional excitation of hydrogen.
With temperatures and densities characteristic of the intergalactic medium close to  high redshift galaxies ($T\sim$ a few$\times 10^4$K, n$\sim 0.1-1$ cm$^{-3}$), collisional excitation of Ly$\alpha$ emission may be an important indicator of gas cooling in galactic potential wells (e.g., Haiman, Spaans, \& Quataert 2000; Dijkstra \& Loeb 2009; Goerdt et al 2010; Faucher-Giguere et a 2010). A significant fraction of the gravitational energy will be carried away by HI 2-photon continuum radiation.

Is it plausible that we may be seeing a combination of Ly$\alpha$ and 2-photon continuum emission in the HST ACS broad-band images ?
To test this possibility we have designed a toy-model to study the expected emission with the Cloudy code (Ferland et al 2013).  The width of the tentacles is spatially unresolved; we take them to be cylindrical, and with radius of R=500 pc and a typical length of L=3.5 kpc. 
For the Cloudy simulation we replace the resulting gas volume by a sphere of equal volume. This is justifiable as we are only testing for
the overall viability of creating the observed continuum emission through collisional excitation, and any geometry will be optically thin
for 2-photon continuum (but optically thick for Ly$\alpha$ photons,  the radiative transfer of which we cannot model anyway).

The gas in the spherical toy model is placed at a constant temperature of 18,000 K, close to the temperature that maximizes collisional excitation of HI. Such a constellation may be found in a
cooling stream where the temperature is kept up by a steady supply of gravitational energy by the infalling matter. A photoionizing UV background according to Haardt \& Madau (1996) is adopted, although the presence  of the extragalactic UV background makes little difference. A metallicity typical for the high redshift IGM (Z=$10^{-2.8}Z_{\odot}$, e.g., Schaye et al 2003, Simcoe et al 2004) was assumed for the gas, but the exact metallicity value is not very important. Even increasing the metallicity in the gas to solar values would increase the  cooling rate by only about 20\%.
 
To determine the cooling flux we fit the fluxes observed within the tentacles' contour in the B, I, z and Y bands (omitting the V band that may contain the Ly$\alpha$ line)
with cooling radiation spectra from cloudy, using the gas density as the only free parameter. A density of 0.75 cm$^{-3}$ gives the best fit, with a reduced $\chi^2_{\nu} = 1.66$ (fig. \ref{tentacont}, central panel). Having determined the strength of the continuum and the intrinsic Ly$\alpha$ flux ($1.3\times10^{-16}$ erg cm$^{-2}$ s$^{-1}$) this way we fit the V band separately, now treating the  fraction of the intrinsic Ly$\alpha$ that is observed within the tentacular contour, $f^{{\rm Ly}\alpha}_{tent}$, as the free parameter. It is found that  18.4\% of the V band flux need to be supplied by Ly$\alpha$ line emission,
corresponding to  $f^{{\rm Ly}\alpha}_{tent}$= $6.2\pm 4.4 \%$. The error comes from the uncertainty in the observed V band flux only and is thus a strict lower limit to the actual uncertainty. Even given the large uncertainty we can say the majority of flux in the V band must come from the 2-photon continuum. 
The small contribution of Ly$\alpha$ to the total emission directly from the volume of the tentacles into the observer's line-of-sight is quite expected, because Ly$\alpha$ is likely
to get scattered out of the tentacles before being able to escape the galactic halo, whereas the 2-photon emission emerges straight from the source into our
line-of-sight.
Under the above conditions, the gas is mostly ionized (neutral fraction 13\%). However, if contained in tentacular tubes with the above dimensions, the gas remains self-shielded, 
with a minimum HI column density  $N_{\rm HI}$ = $1.5\times10^{20}\frac{R}{500{\rm pc}}$cm$^{-2}$. 

The intrinsic Ly$\alpha$ flux ($1.3\times10^{-16}$ erg cm$^{-2}$ s$^{-1}$) from cooling in the filaments overproduces the total observed flux ($2.45\times10^{-17}$ erg cm$^{-2}$ s$^{-1}$, suggesting an overall apparent escape fraction for Ly$\alpha$, $f^{{\rm Ly}\alpha}_{esc}$=0.19, if all the flux in the system is produced through this
channel, or about half that if only the flux in the red core is. Apparent escape probabilities considerably less than unity are not unusual (e.g., Atek et al 2009; Blanc et al 2011) and may be due to destruction of Ly$\alpha$ photons by dust, anisotropic emission, or aperture losses. 

At the optimal temperature of 18,000 K  the tentacle gas, for the adopted volume and density, undergoes collisional cooling at the rate of 
$dE_{cool}/dt=2.3\times10^{43}$erg s$^{-1}$. This is very similar to the  total gravitational heating rate expected for a galaxy with a total mass of $10^{12} M_{\odot}$
at this redshift ($1.6\times10^{43}$erg s$^{-1}$, Goerdt et al 2010; see also Faucher-Giguere et al 2010). 

An indirect estimate of the halo mass can be obtained from the spectral energy distribution. Using the MAGPHYS code (da Cunha, Charlot \& Elbaz 2008)  with
the information obtained from bands ACS F435W, ACS F606W, ACS F775W, ACS F850LP, NICMOS F110W and NICMOS F160W, and allowing for the presence
of dust gives a stellar mass of $2.4\times10^9\ M_{\odot}$, corresponding to a typical halo mass of $\sim 7\times 10^{11} M_{\odot}$ according to theoretical stellar/halo mass relations (e.g., Moster et al 2010). In other words, the total gravitational energy gained by cold accretion  in the halo of the galaxy is of the same
order of magnitude  as the cooling rate required to explain the continuum flux  in the tentacles as HI 2-photon cooling radiation. This result requires
that most the gravitational energy is re-radiated at small radii by the tentacles, which is not implausible. 
In the simulations by Rosdahl \& Blaizot (2012) half of the total cooling energy for halos of similar masses is emitted at radii of less than 10 kpc.  
A possible problem with the interpretation of the tentacular flux as cooling radiation arises from the relatively high star-formation rate suggested by the expected mass accretion rate for a $\sim 7\times 10^{11} M_{\odot}$ halo ($M_{\rm cold}\sim 100\ M_{\odot}$ yr$^{-1}$; e.g. Goerdt et al 2010) and the more measly observed  star-formation rate of 4.1 $M_{\odot}$yr$^{-1}$ (derived from MAGPHYS; this is a somewhat higher value that the 1.7 $M_{\odot}$yr$^{-1}$ we had estimated in paper I, then based solely on the F606W flux before extinction correction). 
If star formation is efficient and most of the accreted gas ends up in stars, $M_{*}\approx M_{\rm cold}$, one would obtain a star formation rate 25 times larger
than the observed 4.1 $M_{\odot}$ yr$^{-1}$).  We are forced to conclude that either star-formation in the present case is currently highly inefficient, or that the
accretion rate and/or the total mass of the galaxy are much smaller. In the latter case, a  galaxy mass commensurate with the observed star-formation rate 
would be $2.5\times10^{10}M_{\odot}$. The corresponding gravitational heating rate for such a halo (Goerdt et al 2010, equation 7) would then be almost
three orders of magnitude smaller than required to produce the observed cooling continuum. Of course, any of those arguments are applicable only in a statistical sense.

\subsection{Fluorescent recombination emission}

Alternatively, nebular continuum emission can arise when HI gas fluoresces in response to ionizing radiation from either stars or an AGN
(for discussions see,  e.g., Cantalupo et al 2005; Kollmeier et al 2010). Both the energetics and the shape of the emerging spectrum depend on a considerable number of parameters,
including the distance of the gas from the sources, the column density along the line of sight and between source and irradiated gas cloud, the ionizing spectrum,
the relative prominence of collisional versus photoionization excitation, and the metallicity of the gas, so we can at best check the plausibility of this process.
The result of one particular realization is shown in the bottom panel of fig.\ref{tentacont}.
Using Cloudy  we model the "reflected" fluorescent radiation when bouncing into our line-of-sight off a gas cloud irradiated
by a $5\times10^6$yr old starburst. Again omitting the V band from the fit (because of its susceptibility to the Ly$\alpha$ escape fraction), we obtain a fit
to the observed continuum colours in the tentacles with $\chi^2_{\nu} = 0.85$, as good as the fit with the stellar continuum above. The somewhat better agreement
for the fluorescent case as compared to the earlier collision excitation case is owed to the production of diffuse HI ionizing continuum emission (which was absent in the case of collisional excitation).
However, the relative strength of the actually observed nebular ionizing continuum may depend strongly on the position of the gas clouds relative to the ionizing source and the observer, so the better fit for the fluorescent case may be fortuitous.
Other differences between the fluorescent spectrum and the mainly 2-photon continuum of the collisional case are the presence of metal recombination lines (which are
weak here as we assumed the mean metallicity of the IGM at $z\sim 3$, $Z=10^{-2.8} \times Z_{\odot}$ (e.g., Schaye et al 2003, Simcoe et al 2004), and a shallower continuum slope in the red due to the fluorescent Balmer continuum. 

Explaining the tentacular continuum as fluorescent would be consistent with our earlier proposal that the extended Ly$\alpha$ emission itself is due to fluorescence
induced by ionizing stellar photons that have escaped the galaxy. The absence of a detectable underlying stellar population within the footprint of the Ly$\alpha$ filament 
and the lack of surface brightness gradients expected if Ly$\alpha$ photons were "random-walking" outward from an optically thick galaxy
make this the most plausible scenario.

We cannot rule out the presence of an AGN, which would be expected to leak photons below the Lyman limit and could lead to similar, large scale Ly$\alpha$ emission,
but, as argued earlier,  the number of ionizing photons produced by the observed stars is sufficient to explain the observed Ly$\alpha$ flux, and the peculiar morphology of the
galaxy certainly suggests that star formation is the relevant process here. 

\section{Outflows versus Inflows - physical nature of the tentacles}

The astrophysical nature of the tentacles cannot be decided just from knowing the origin of the tentacles emission (i.e., stellar versus gas cooling vs. photoionization), and
additional morphological or kinematic information is required.

We shall briefly discuss four scenarios that may be capable of producing galaxies with tentacles as observed:

(1) galactic winds; (2) ram pressure stripping; (3)  galactic wakes forming behind satellite halos when passing through the main halo; and (4)  cold accretion streams connecting to the galaxy.  

(1): Even though the large-sale filament may be in-falling gas, this does not have to be the case
for the smaller tentacles, which have some morphological resemblance to gaseous filaments seen in galactic winds (e.g., Veilleux et al 2005). 
Although the total star formation rate of 4.1 $M_{\odot}$yr$^{-1}$ is moderate, the star formation rate density should be enough to drive a galaxy wide wind (according to the
criterion given by Heckman 2002). Somewhat surprisingly for an outflow, the tentacles appear strictly on one side, and not in the bi-polar pattern expected. Besides they seem to converge on or emerge from a location slightly in front of the galaxy (in the direction of the Ly$\alpha$ emission) that does not agree with the luminosity peak (and presumed peak
of star formation). If the tentacles were mainly illuminated by fluorescence, such an asymmetric pattern with a wind cone visible only on one side would be expected (with the other cone present but not illuminated), and recombination could be sufficient to produce the observed continuum. Cooling radiation from dense, cold, entrained matter appears less likely as it would arise equally strong in both wind cones. Star formation triggered by some types of outflows, e.g.,  galactic superbubbles (Oey et al 2005) and AGN outflows (e.g., Van Breughel et al 1985; Rauch et al 2013b)
has been observed, but the filamentary structures sometimes seen in galactic winds do not seem to be populated with newly formed stars, so a tentacular structure of stars may be an unlikely outcome for a wind scenario. Finally, as can be seen from the spectral cuts in fig.\ref{cuts}, beyond
the innermost few kpc the Ly$\alpha$ emission filament remains more or less perpendicular to the dispersion direction, which means
that the gas is neither accelerating nor decelerating, as may be expected for various wind models. On the other hand, velocity gradients would be hard to see if the wind
and the orientation of the tentacles were largely perpendicular to the line-of-sight. In short, the possibility that the tentacles arise in a galactic wind
cannot be ruled out, but it is not obvious either. 

(2) Ram pressure stripping may create linear filaments of new star formation downwind from the stripped galaxy, which would indeed explain the occurrence of emitting tentacles strictly
on only one side of the galaxy. However, the stripped contrails observed so far at lower redshifts are generally parallel to each other, following hydrodynamic streamlines away from the galaxy (e.g., 
Hester et al 2010; Ebeling et al 2014), in contrast to the the highly focused  bundle of tenacles observed in the present case. Thus, ram-pressure stripping
of the galaxy itself when moving with respect to the surrounding intergalactic medium appears an unlikely explanation.

(3)  Accretion streams in the cosmic web consist not just of gas but of galactic halos as well and must be exposing the main galaxy to a constant bombardment with
satellite halos. These objects, as far as they are not getting completely destroyed on their passage through the galaxy, will be tidally and pressure-stripped and may re-emerge post-passage with a new tail
of ripped-out interstellar medium from the galaxy, with trajectories bent by the gravitational potential of the main galaxy. These processes are seen in cosmological simulations (e.g., Agertz et al 2010, Rosdahl \& Blaizot 2012; Keres et al 2012). While the occurrence of satellite tails
is difficult to observe at the level of individual high redshift galaxies,  they must be common, and play an important role in the metal-enrichment of galactic halos at high z. The 'tail' to the West (right in fig. \ref{blowup}) of the galaxy is almost certainly of this kind. The tentacles look very similar to multiple tails passing through the lowest mass halo shown by Rosdahl \& Blaizot (2012) in their figure 10. The similar morphology, however, does not necessarily require that the emission mechanisms are the same.
The detectable tentacular emission in the Rosdahl \& Blaizot simulated halo appears dominated by gravitational cooling radiation, 
but in our case the tentacles could just as well be illuminated by fluorescence from the galaxy.

(4) The tentacles may be actual cold accretion, i.e., inflowing gas conduits connecting the large scale accretion filament to the
ISM of the galaxy. When the relatively wide (kpc scale) IGM accretion streams converge on a galaxy they must get pinched to a much smaller diameter by the ambient pressure of the galactic halo. This compression and shrinking of the in-falling filaments is seen in some simulations (e.g., Agertz et al 2010). We can speculate that
the compression may enhance cooling radiation and may even induce star-formation in the gas while in final descent onto the galaxy. If these processes take place,
they may lead to the formation of one or several bright knots of emission at the convergence of the tentacles, and may explain the general
lining up of these  tentacles with the direction of the large scale filament that we see in Ly$\alpha$ emission, and the large scale distribution of galaxies further out.

\section{Conclusions}

We have discussed the nature of an asymmetric Ly$\alpha$ emitter associated with a peculiar star-forming galaxy. As described in paper I, spatial distribution and kinematics of the
Ly$\alpha$ emitting gas suggest a picture of gas falling into a galaxy while being lit up by ionizing radiation escaping from that object.
Based on an examination of HST imaging data, the underlying galaxy exhibits a "tadpole" shape in the V band, with a hot spot of blue continuum emission on the edge facing the direction of the massive Ly$\alpha$ emission. The galaxy further shows a stretch of continuum light likely to represent a tidally or pressure-stripped passing satellite galaxy, and several sharply delineated  tentacles of continuum emission protruding to about 5 kpc from the galaxy in the direction toward the Ly$\alpha$ emission. 
The morphology of the tentacles and their similar spatial range and surface brightness suggest that we are seeing extragalactic nebular emission dominated by HI 2-photon continuum emission (as opposed to star light). The broad band colours and the energy requirements are consistent with collisionally excited cooling radiation powered by potential energy
of gas falling into a galaxy of this type. However, the observed current star formation rate of the galaxy falls short of matching the predicted mass accretion rate by a factor $\sim 25$. Perhaps more compellingly, we may be seeing recombination radiation from gas fluorescing in response to ionizing radiation escaping from the galaxy. As argued in paper I, this
explanation is energetically compatible with the observed star formation. It would also most easily explain the presence of Ly$\alpha$ emission in the large-scale filament, tens of kpc away from the galaxy as being lit up directly by a beam of escaping ionizing radiation. Alternatively, in the absence of any discernable local sources residing in the Ly$\alpha$ emitting region, 
its Ly$\alpha$ would have to be either propagated by scattering through an optically thick halo all the way from the galaxy, leading to an (unobserved) strong gradient in 
the surface brightness as a function of distance, or be produced by cooling at distances from the galaxy as large as several tens of kpc, where the density is likely to be lower than the $\sim 1$ cm$^{-3}$ required to produce detectable cooling radiation in the current case.

The emission mechanism alone (nebular emission, either collisional or recombination continua) is insufficient to uniquely identify the astrophysical setting of the tentacular  
phenomenon. The tentacles could be features of a galactic wind blown by star formation that is triggered by the in-falling gas. Alternatively, theoretical infall scenarios
predict cold stream filaments converging at galaxies which look very similar to the observed structures. A third explanation, within the same accretion scenario, might envisage
in-falling dwarf satellite halos piercing the interstellar medium of the main halo and pulling out dense filaments of gas, which could be seen either while cooling or again
when being exposed to ionizing photons.

In any case, the combined evidence from the elongated gas distribution, the kinematics of the gas, the large scale environment, the apparent hot spot of star-formation,
the nebular continuum emission, the likely escape of ionizing photons, and the presence of at least one interacting satellite halo appears to be intriguing evidence
for an instance of a galaxy undergoing accretion of gas and smaller halos from the intergalactic medium. Until it becomes instrumentally
possible to detect the Ly$\alpha$ glow of the IGM in response to the UV background, the search for galaxies that illuminate
themselves through some fortuitous release of Ly$\alpha$ or ionizing radiation into their environment may provide our main direct insights
into the in- and outflows of gas. Searches for asymmetric Ly$\alpha$ halos or offsets between stellar populations and Ly$\alpha$ emission may reveal further objects where the escape of ionizing radiation can be studied. 

The detection of extragalactic 2-photon continuum emission in space-based, broad-band images of galaxies flagged by
asymmetric Ly$\alpha$ emission may become  a new tool for probing the the gaseous environment of high redshift galaxies. 
Unlike Ly$\alpha$ or other resonance lines, the 2-photon continuum emission is not spatially “blurred” as it is emitted under optically thin conditions, suggesting that it can be used to “image” small-scale features in the IGM at much greater spatial resolution. 
\smallskip

\section*{Acknowledgments}

We thank the the staff of Las Campanas Observatory for their help with the observations.
This paper is based partly on observations made with the NASA/ESA Hubble Space Telescope, and obtained from the Hubble Legacy Archive, which is a collaboration between the Space Telescope Science Institute (STScI/NASA), the Space Telescope European Coordinating Facility (ST-ECF/ESA) and the Canadian Astronomy Data Centre (CADC/NRC/CSA).  This research has made use of the NASA/IPAC Extragalactic Database (NED) which is operated by the Jet Propulsion Laboratory, California Institute of Technology, under contract with the National Aeronautics and Space Administration. 
MR was supported by grant AST-1108815 from the National Science Foundation. This work was supported by the ERC Advanced Grant 320596 
“The Emergence of Structure during the epoch of Reionisation”.

\end{document}